\documentclass{aa}
\usepackage[varg]{txfonts}
\usepackage{graphicx}
\graphicspath{{figures/}}
\usepackage{natbib}
\bibpunct{(}{)}{;}{a}{}{,} 
\usepackage{hyperref}
\hypersetup{
        colorlinks=true,
        linkcolor=blue,
        citecolor=blue
}

\begin{document}

\title{Halo shapes constrained from a pure sample of central galaxies in KiDS-1000}

\author{Christos Georgiou \inst{1}\thanks{georgiou@strw.leidenuniv.nl}
        \and Henk Hoekstra \inst{1}
        \and Konrad Kuijken \inst{1}
        \and Maciej Bilicki \inst{2}
        \and Andrej Dvornik \inst{3}
        \and Thomas Erben       \inst{4}
        \and Benjamin Giblin \inst{5}
        \and Catherine Heymans \inst{3,5}
        \and Hendrik Hildebrandt \inst{3}
        \and Jelte T. A. de Jong        \inst{6}
        \and Arun Kannawadi     \inst{1,7}
        \and Peter Schneider \inst{4}
        \and Tim Schrabback \inst{4}
        \and HuanYuan Shan \inst{8,9}
        \and Angus H. Wright \inst{3}
}
\institute{Leiden Observatory, Leiden University, Niels Bohrweg 2, 2333 CA, Leiden, The Netherlands.
        \and Center for Theoretical Physics, Polish Academy of Sciences, al. Lotnik\'ow 32/46, 02-668, Warsaw, Poland.
        \and Ruhr University Bochum, Faculty of Physics and Astronomy, Astronomical Institute (AIRUB), German Centre for Cosmological Lensing, 44780 Bochum, Germany
    \and Argelander-Institut f\"ur Astronomie, Universit\"at Bonn, Auf dem H\"ugel 71, 53121, Bonn, Germany.
        \and Institute for Astronomy, University of Edinburgh, Royal Observatory, Blackford Hill, Edinburgh, EH9 3HJ, UK
        \and Kapteyn Astronomical Institute, University of Groningen, PO Box 800, 9700 AV Groningen, the Netherlands
        \and Department of Astrophysical Sciences, Princeton University, 4 Ivy Lane, Princeton, NJ 08544, USA
        \and Shanghai Astronomical Observatory (SHAO), Nandan Road 80, Shanghai 200030, China
        \and University of Chinese Academy of Sciences, Beijing 100049, China}
        
\date{Received <date> / Accepted <date>}

\abstract{
We present measurements of $f_h$, the ratio of the aligned components of the projected halo and galaxy ellipticities, for a sample of central galaxies using weak gravitational lensing data from the Kilo-Degree Survey (KiDS). Using a lens galaxy shape estimation that is more sensitive to outer galaxy regions, we find $f_{\rm h}=0.50\pm0.20$ for our full sample and $f_{\rm h}=0.55\pm0.19$ for an intrinsically red sub-sample (that therefore has a higher stellar mass), rejecting the hypothesis that round halos and/or galaxies are not aligned with their parent halo at $2.5\sigma$ and $2.9\sigma$, respectively. We quantify the 93.4\% purity of our central galaxy sample using numerical simulations and overlapping spectroscopy from the Galaxy and Mass Assembly survey. This purity ensures that the interpretation of our measurements is not complicated by the presence of a significant fraction of satellite galaxies. Restricting our central galaxy ellipticity measurement to the inner isophotes, we find $f_{\rm h}=0.34\pm0.17$ for our red sub-sample, suggesting that the outer galaxy regions are more aligned with their dark matter halos than the inner regions. Our results are in agreement with previous studies and suggest that lower mass halos are rounder and/or less aligned with their host galaxy than samples of more massive galaxies, studied in galaxy groups and clusters.}


\maketitle

\section{Introduction}
\label{sec:Introduction}

The current standard model of cosmology, dubbed $\Lambda$CDM, has been very successful in describing a large number of independent cosmological probes, such as cosmic microwave background (CMB) observations \citep[e.g.][]{CMB}, the galaxy clustering signal \citep[e.g.][]{clustering}, and baryon acoustic oscillations \citep[e.g.][]{BAO1,BAO2}, among many others. According to this model, dark matter makes up the majority of the matter density content of the Universe and provides the seeds upon which galaxies and larger structures can form and evolve.

From numerical simulations, it is understood that dark matter forms halos that are roughly tri-axial; they appear elliptical in projection \citep{Triaxial1, Triaxial2}. An estimation of the shape of these halos from observations can therefore be used as a test for the current cosmological model, as well as extensions to it, such as modifications to the gravity theory or the dark matter component \citep[e.g.][]{ModifiedgravityHalo1, ModifiedgravityHalo2, SIDMHalo, WDMHalo}.

Observationally, many attempts have been made to measuring halo ellipticities. Techniques include satellite dynamics \citep[e.g.][]{Satellites1,Satellites2,Satellites3,Satellites4}, tidal streams in the Milky Way \citep[e.g.][]{MWstream1,MWstream2,MWstream3}, HI gas observations \citep[e.g.][]{HI0,HI1,HI2}, planetary nebulae \citep[e.g.][]{PN1,PN2}, X-ray observations \citep[e.g.][]{Xray}, and strong lensing \citep[e.g.][]{StrongLens}, also accompanied by stellar dynamics \citep[e.g.][]{StDynStLens1}. These techniques rely on luminous tracers of the dark matter shape, which can lead to biases, complicate the interpretation of the measurements, and cannot provide information on the larger scales of the dark matter halo where visible light is absent. 

One observational technique without this drawback is weak gravitational lensing, the coherent distortion of light rays from background sources from the intervening matter distribution \citep[for a review, see][]{WLreview}. Because gravitational lensing is sensitive to all matter (also non-baryonic matter), it serves as a great tool for studying the dark matter halos. The distortion of the galaxy shapes due to weak lensing is very small, and in order to extract a measurable signal, we need to statistically average over large ensembles of galaxies. When the stacking is made around other galaxies, a technique called galaxy-galaxy lensing, only the matter at the lens galaxy redshift contributes coherently to the lensing signal, and the structures along the line of sight simply add noise to the measurement. 

Tri-axial dark matter halos cause an azimuthal variation in the weak-lensing signal, enhancing it along the direction of the semi-major axis of the projected halo and reducing it along the semi-minor axis. For very massive structures, such as large galaxy clusters, this variation is strong enough to be measured for individual \citep[e.g.][]{ClustersWL,Clash} or stacked weak-lensing maps of cluster samples \citep{Evans&Bridle, Oguri2012}. For galaxy-scale halos, this variation can be measured by weighting the lensing measurements according to the halo semi-major axis. 

In most applications of weak-lensing-based measurements of the dark matter halo ellipticity, the lens galaxy semi-major axis is used as a proxy for the dark matter halo axis \citep{Hoekstra2004,Mandelbaum2006,Parker,Edo2012,Tim2015,Edo2017, Tim2020}. The measured quantity is then the ratio of the halo ellipticity to the galaxy ellipticity, weighted by the average misalignment angle between the two, that is, $f_{\rm h}=\langle\cos(2\Delta\phi_{\rm h,g}) |\epsilon_{\rm h}|/|\epsilon_{\rm g}|\rangle$. This makes the measurement of $f_{\rm h}$ a useful step in determining the alignment between the dark matter halo and its host galaxy. The misalignment angle has been measured in numerical simulations, with results from the most recent hydrodynamical simulations suggesting a value of $\langle\Delta\phi_{\rm h,g}\rangle\sim30^\circ$ \citep{Tenneti2014,Marco2015,Elisa2017}. The misalignment decreases with decreasing redshift and increasing halo mass, which suggests that massive central galaxies are expected to carry most of the signal. \citet{Edo2017} detected a non-zero halo ellipticity with $\gtrsim3\sigma$ significance using only $\sim2500$ lenses. These lenses were confirmed central galaxies of a galaxy group from a friends-of-friends-based group catalogue 
built using spectroscopic data.

Motivated by this, we aim to define a sample of central galaxies with very high purity from a \emph{\textup{photometric}} galaxy sample and use these as lenses to measure the anisotropic weak-lensing signal around them. A galaxy sample with low satellite fraction also produces a more robust measurement because satellite galaxy lensing profiles across a wide range of scales complicate the interpretation of the measured signal of the full sample. We used the fourth data release of the Kilo-Degree-Survey \citep[KiDS,][]{KIDSDR4} and constructed an algorithm that preferentially selects central galaxies using apparent magnitudes and photometric redshifts. These redshifts are obtained from a machine-learning technique, focussing on the bright-end sample of galaxies in KiDS, and they achieve very high precision \citep{Maciek}. We validated our central galaxy selection by quantifying the sample purity using the group catalogue from the Galaxy And Mass Assembly survey \citep[GAMA,][]{Driver2011,Aaron} and mock galaxy catalogues from the Marenostrum Institut de Ci\'encies de l’Espai (MICE) Grand Challenge run \citep{MICE2b}. 

In Sect. \ref{sec:Data} we present the data used for constructing and validating our lens sample, consisting of highly pure central galaxies, which we describe in detail in Sect. \ref{sec:Centrals}. The method used to measure the lensing signal is described in Sect. \ref{sec:Methodology}. The results obtained are shown in Sect. \ref{sec:Ellipticity}, and we discuss the measurements and conclude in Sect. \ref{sec:Conclusions}. To calculate angular diameter distances, we use a flat $\Lambda$CDM cosmology with parameters obtained from the latest CMB constraints \citep{CMB}, that is, $H_0=67.4$ km$/$s$/$Mpc and $\Omega_{m,0}=0.313$.

\section{Data}
\label{sec:Data}

Measuring the anisotropic lensing signal requires a wide survey of deep imaging data so that accurate unbiased galaxy shapes can be measured and the lensing signal can be statistically extracted. For this reason, we used data from KiDS. Moreover, massive central galaxies are expected to yield the highest signal-to-noise ratio (S/N) for anisotropic lensing; we thus need a way of selecting a pure sample of central galaxies as well as a means to validate our selection. To this end, we made use of the GAMA survey and mock catalogues from the MICE Grand Challenge galaxy catalogue.

\subsection{KiDS-1000}
\label{sec:KiDS}

KiDS\footnote{http://kids.strw.leidenuniv.nl} \citep{deJong2015, deJong2017, KIDSDR4} is a deep-imaging ESO public survey carried out using the Very Large Telescope (VLT) Survey Telescope and the OmegaCam camera. The survey has covered 1,350 deg$^2$ of the sky in three patches in the northern and southern equatorial hemispheres in four broad-band filters ($u$, $g$, $r,$ and $i$). The mean limiting magnitudes are 24.23, 25.12, 25.02, and 23.68 for the four filters (5$\sigma$ in a 2$''$ aperture). The survey was specifically designed for weak-lensing science and the image quality is high, with a small nearly round point-spread function (PSF), especially in the $r$-band observations, which were taken during dark time with the best seeing conditions. We used the fourth data release of the survey, with 1006 $1\times1$ deg$^2$ image tiles (KiDS-1000).

KiDS is complemented by the VISTA Kilo-Degree Infrared Galaxy Survey \citep[VIKING,][]{VIKING}, which has imaged the same footprint as KiDS in the near-infrared (NIR) $Z,Y,J,H,$ and $K_s$ bands. This addition allows determining more accurate photometric redshifts from nine broad-band filters. For our source galaxy sample, redshifts were retrieved with the template-fitting Bayesian photometric redshift (BPZ) code \citep{BPZ1,BPZ2}, applied to the nine-band photometry. To estimate source redshift distributions, we used the direct calibration scheme to weight the overlapping spectroscopic sample according to our photometric one. The process is described in detail in \citet{KV450}. 

For our lens sample, we required more precise redshift estimates that help in a more accurate lensing measurement, as well as to build a more robust central galaxy sample. We therefore chose to use a bright ($m_r\lesssim20$) sample with photometric redshifts estimated with the artificial neural network machine-learning code ANNz2 \citep{ANNZ}, as presented in \citet{Maciek}, but now extended to the full KiDS-1000 sample \citep{Bilicki}. This sample was trained on the highly complete GAMA spectroscopic redshift catalogue ($98.5\%$ completeness at flux limit $m_r<19.8$ in equatorial fields, \citealt{Liske2015}). The full overlap between GAMA equatorial and KiDS and the unbiased selection of flux-limited spectroscopy in the former dataset allowed us to obtain very precise and accurate photometric redshift estimates for our lens sample, with a mean bias $\langle\delta z\rangle = \langle z_\mathrm{phot} - z_\mathrm{spec}\rangle \simeq 10^{-4}$ and scatter $\sigma_{\delta z} \simeq 0.02(1+z)$. By adding VIKING data over the full KiDS-1000 area \citep{KIDSDR4}, the default photo-$z$ solution is now based on nine-band photometry.
In this work, however, we used redshifts obtained from the optical $ugri$ band photometry alone: because the lens sample is bright and has a relatively low redshift, NIR photometry does not significantly improve the photometric redshift estimation (see \citealt{Maciek} for more details). In addition, using the NIR photometry would introduce additional masking to our data because the VIKING coverage has some gaps, which would reduce our lens galaxy sample. As we show in Sect. \ref{sec:Centrals}, improving the redshift accuracy further (e.g. with NIR data) does not significantly increase the purity of our central sample, whereas increasing the survey area equips us with a larger sample for a more precise measurement. We restricted the lens redshifts to $0.1<z_{\rm l}<0.5$ because outside of this range the photo-$z$s are less well constrained \citep{Maciek}; this cut removes a small fraction of the lens sample ($<10\%$).

Galaxy shapes for our source galaxy sample were measured using the THELI\footnote{https://www.astro.uni-bonn.de/theli/}-reduced $r$-band images with the \emph{lens}fit shape measurement method \citep{lensfit1,lensfit2, Giblin}. This method is a likelihood-based algorithm that fits surface brightness profiles to observed galaxy images and takes the convolution with the PSF into account. Within a self-calibrating scheme, it has been shown to measure galaxy shear to percent-level accuracy in simulated KiDS $r$-band images \citep{FC,Arun}. 

Shears were obtained using \emph{lens}fit for galaxies with an $r$-band magnitude larger than 20, which does not allow us to use them for the shapes of our lens sample. In addition, \emph{lens}fit is optimised for galaxies with low S/N and no set-up for measuring bright galaxies is readily available. To acquire shape information for our lens sample, we applied the \textsc{DEIMOS} shape measurement method \citep{DEIMOS} to the \textsc{AstroWise}\footnote{http://www.astro-wise.org/} reduced $r$-band KiDS images. 

\textsc{DEIMOS} is based on measuring-weighted surface brightness moments from galaxy images and using these to infer the galaxy ellipticity. Unlike other moment-based techniques, it allows for a mathematically accurate correction of the PSF convolution with the galaxy light profile, avoiding any assumptions on the profile or PSF behaviour. The accuracy of this correction is only limited by the accuracy of the PSF modelling. Moreover, a correction for the necessary radial weighting, employed during moment measurement, is used; higher-order moments are calculated in order to approximate the unweighted galaxy moments from measured weighted moments.

To model the PSF, we used shapelets \citep{shapelets}. These are orthogonal Hermite polynomials multiplied with Gaussian functions that can be linearly combined to describe image shapes. The process is described in \citet{Kuijken}, who reported that the model performed very well in KiDS imaging data and displayed only a very weak residual correlation between the modelled ellipticities and those measured using the stars in the image. To measure galaxy moments, we used an elliptical Gaussian weight function following a per-galaxy matching procedure. The size of the weight function is tied to the scale of this Gaussian, and we used two different scales in this work, equal to the isophote of the galaxies $r_{\rm iso}$ and $1.5r_{\rm iso}$ (defined at 3$\sigma$ above the background, see \citealt{Georgiou1} for more details). We used these two values to probe potential differences in the measured ellipticity ratio with the galaxy scale probed; a larger weight function will reveal more of the shape of the outer galaxy regions. Neighbouring sources in the image were masked using segmentation maps from \textsc{SExtractor} \citep{sextractor}. A detailed description of the shape measurement process can be found in \citet{Georgiou1}. 

For the GAMA galaxy sample, which is very similar in properties to the lens sample used here, \citet{Georgiou1} showed that the multiplicative bias on the ellipticity (not shear) is lower than 1\% and does not depend strongly on the galaxy properties. This is attributed to the great flexibility of the \textsc{DEIMOS} method and to the fact that these galaxies have a very high S/N in the KiDS imaging data (with a mean S/N $\sim$ 300 in $r$-band images) and are generally very well resolved compared to the PSF size.

In our analysis, we did not probe the lensing signal on very large scales and therefore did not subtract the signal around random points, which in any case has been shown to be consistent with zero in other KiDS weak-lensing measurements \citep{Dvornik}. Additive bias in the shape measurements is not expected to bias the spherically averaged gravitational shear measurements. The anisotropic lensing measurements are not expected to be affected either because sources and lenses were measured using different shape measurement methods, and any additive biases (which are measured to be negligibly small) are not expected to be correlated. Multiplicative biases are also not expected to play a significant role because they affect the isotropic and anisotropic lensing signal in the same way, which would leave the measurement of halo ellipticity unaffected. Furthermore, multiplicative bias for the lens shapes has been shown to be on the sub-percent level \citep{Georgiou1} and does not affect the calculation of the position angle of the lens.

\subsection{GAMA}

GAMA\footnote{http://www.gama-survey.org/} \citep[][]{Driver2009,Driver2011,Liske2015} is a spectroscopic survey carried out with the Anglo-Australian Telescope, using the AAOmega multi-object spectrograph. It provides spectroscopic information for $\sim300,000$ galaxies over five sky patches of $\sim$ 60 deg$^2$ area each for a total coverage of $\sim286$ deg$^2$. The three equatorial patches (G09, G12, and G15) have a completeness of 98.5\% and are flux limited to $r_{\rm petrosian}<19.8$ mag; of the two south patches, G23 overlaps the KiDS footprint and has a completeness of 94.5\% with a flux limit $i<19.2$ mag. This latter selection gives a slightly lower mean redshift than in the equatorial fields, therefore we used for the lens photo-$z$ training only the deeper and more complete equatorial data. We verified that adding G23 does not improve the photo-$z$ estimates \citep[see][for more details]{Bilicki}. 

\begin{figure}
        \resizebox{\hsize}{!}{\includegraphics{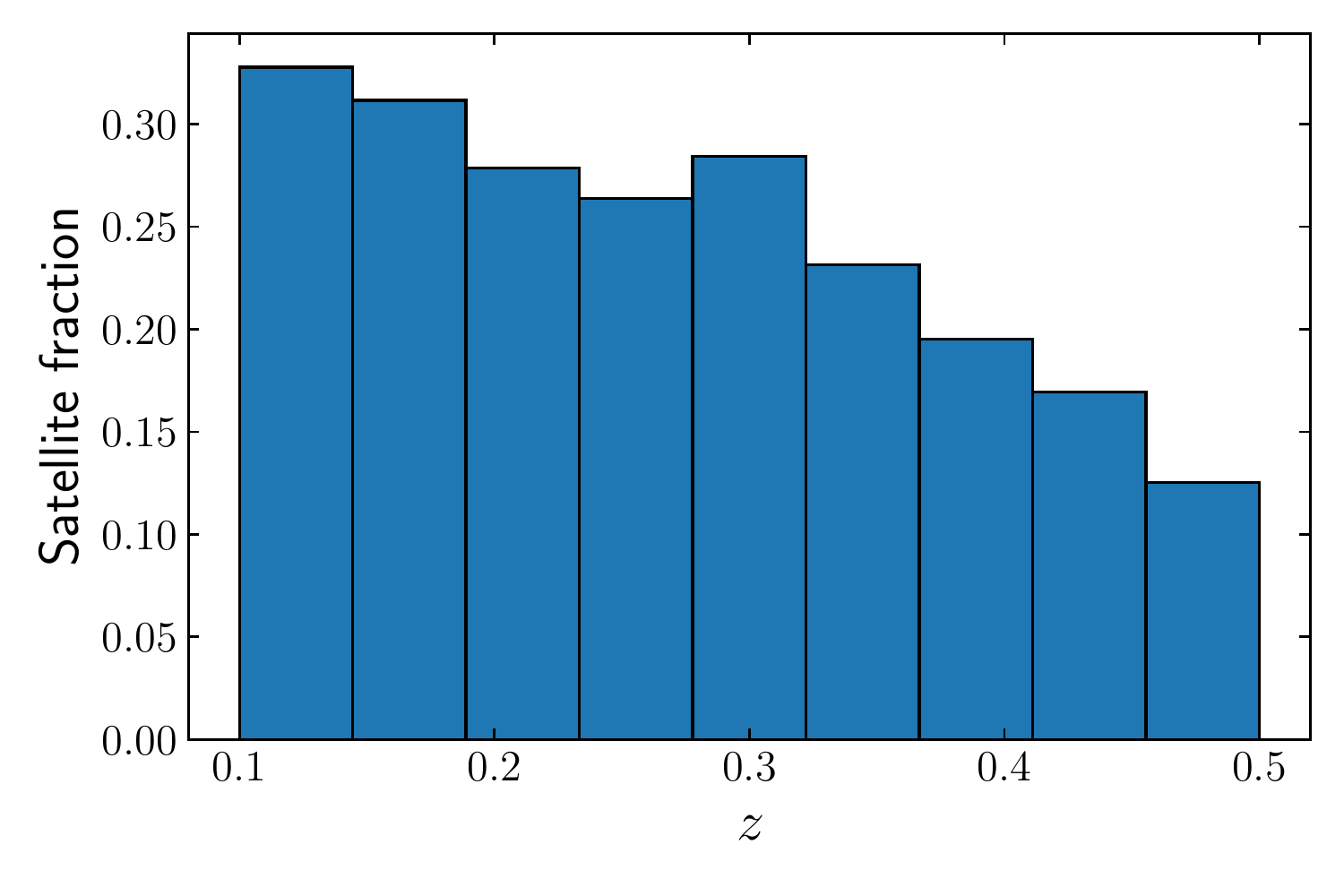}}
        \caption{Satellite fraction ($N_{\rm sat}/N_{\rm all}$) in the GAMA galaxy sample in bins of redshift.}
        \label{fig:satfrac}
\end{figure}

The unique aspect of the GAMA sample is the high completeness, together with the fact that no pre-selection is made on the target galaxies except for imposing a flux limit and removing stars and point-like quasars \citep{Baldry2010}. This nullifies any selection effects and provides the means to produce a highly pure and accurate group catalogue \citep{Aaron}. This catalogue is produced using a friends-of-friends based-algorithm to define galaxy groups and assign galaxies to them. We used this group catalogue to validate our central galaxy sample selection from our lens galaxy sample and quantify its purity, assuming the satellites identified in the catalogue to be the true satellites of the sample. We use the tenth version of this group catalogue, which does not contain the G23 region. 
After masking the lens sample according to the KiDS mask, we are left with $\sim120,000$ galaxies that match the group galaxy catalogue.  In Fig. \ref{fig:satfrac} we show the satellite fraction of the sample in redshift bins; at higher redshift, satellites fall below the detection limit and the satellite fraction of the sample decreases.

\subsection{MICE}
\label{sec:DataMice}

The GAMA group catalogue we used has imperfections, especially for the more massive groups. \citet{Aaron} showed that the number of high-richness groups was lower than expected from mock group catalogues that were specifically designed to validate the group-finding algorithm. In addition, \citet{Arthur}  used hydrodynamical simulations and found that the group algorithm tends to fragment larger groups into smaller ones. Because of this, we chose to also validate our central sample selection using mock galaxy catalogues from a cosmological simulation, the MICE Grand Challenge run \citep{MICE2b}.

MICE is an N-body simulation containing $\sim70\times10^{10}$ dark matter particles in a $(3h^{-1}$Gpc$)^3$ comoving volume, from which a mock galaxy catalogue was built using halo occupation distribution and abundance matching techniques \citep{MICE2d}. Halos are resolved down to a few $10^{11}$ M$_\odot/h$. The catalogue contains information for a large number of galaxy properties, such as apparent magnitude, stellar mass and a distinction of the galaxies into centrals and satellites, which we used here. Other applications of the catalogue include galaxy clustering, weak lensing, and higher-order statistics \citep{MICE2a,MICE2c,MICE2e}. We downloaded the publicly available version 2 of the catalogue from cosmohub\footnote{https://cosmohub.pic.es} \citep{Cosmohub}. From the 5000 deg$^2$ that the whole mock catalogue covers, we cut out 200 deg$^2$ and selected galaxies with apparent SDSS-like $r$-band magnitude of $<20.3$ mag to match the cut performed in \citet{Maciek}.

\section{Central galaxy sample}
\label{sec:Centrals}

\subsection{Algorithm}

In order to optimally extract the anisotropic weak-lensing signal of elliptical dark matter halos, it is important to exclude galaxies in our sample that reside in sub-halos, that is, satellite galaxies \citep[see e.g.][]{Edo2017}. Because of the hierarchical structure formation, central galaxies are commonly found in overdense regions of the Universe where other neighbouring galaxies are also likely to be found. Based on this, we developed an algorithm to search for galaxies in our sample that have a high probability of being a central halo galaxy. 

The algorithm works as follows: For every galaxy in our sample, we searched for neighbouring galaxies inside a cylinder in sky and redshift space. The cylinder radius has a fixed physical length while the depth of the cylinder is determined by the accuracy of our redshift estimation. If neighbouring galaxies are indeed found, we ask whether the galaxy we selected that lies in the middle of the cylinder is the brightest galaxy (in the $r$ band) inside that cylinder. If this is true, we identify this galaxy as a central. We tested two different cylinder depths, $\pm\mathrm{d}z$ and $\pm2\mathrm{d}z$ (where $\mathrm{d}z$ is the redshift uncertainty, equal to $\sim0.02(1+z)$ for our lens galaxy sample) and chose the latter, which was found to perform better.

\subsection{Sample purity}

We tested the performance of this algorithm on the GAMA galaxy survey sample and on the mock galaxy catalogues from the MICE simulation. The spectroscopic information together with the high completeness of the GAMA sample allows constructing a highly accurate group galaxy catalogue, which we used here to identify central and satellite galaxies. We selected central galaxies by removing any galaxy that is a satellite (we kept the brightest group galaxies as well as field galaxies; the latter are expected to live in their own isolated dark matter halo or have satellites around them that are too faint to detect). 

However, this group catalogue is not perfect (see Sect. \ref{sec:DataMice}). Therefore we also used the MICE mock galaxy catalogues to validate our algorithm, where we know a priori the central and satellite galaxies. We mimicked the photometric redshift uncertainty in the mock catalogue redshifts by adding a random number to them, drawn from a Gaussian distribution with a scale equal to the redshift uncertainty,  $\sim0.02(1+z)$.

\begin{figure}
        \resizebox{\hsize}{!}{\includegraphics{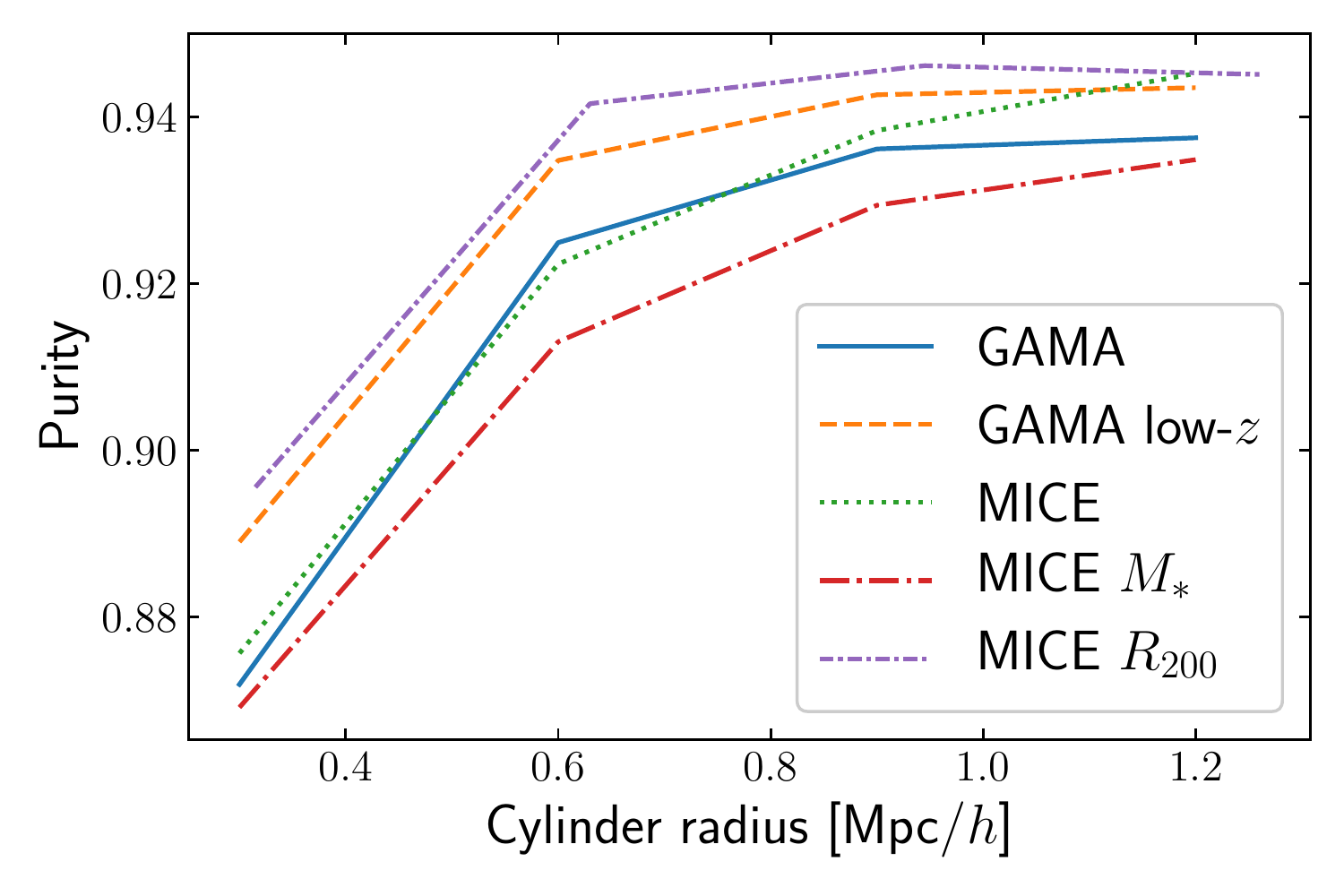}}
        \caption{Purity of our central galaxy sample as a function of fixed cylinder radius used to identify centrals in overdense regions. Lines connect the individual points. In solid blue we show results from the GAMA+KiDS-1000 overlap in the photometric redshift space of $0.1<z<0.5$. We also show the results for redshifts between $0.1<z<0.3$ with a dashed orange line. The dotted green line shows the purity of the sample obtained using the MICE2 mock catalogues for $0.1<z<0.5$. The dash-dotted red line shows results obtained when we searched for the most massive (in terms of stellar mass) galaxy in the cylinder centre instead of the brightest one. Finally, the  dense dash-dotted purple line represents results obtained when instead of a fixed cylinder radius we used multiples $k$ of the galaxy $R_{200}$ to define the radius size, with $k=\{1,2,3,4\}$. In this case, we plot the median value of the cylinder radius on the $x$-axis, corresponding to the four different purity values.}
        \label{fig:purcom}
\end{figure}

We show the performance of our algorithm in Fig. \ref{fig:purcom}, where we plot the purity (number of true centrals we identified over the total number of centrals we identified) of our central galaxy sample as a function of the fixed cylinder radius used. When the GAMA survey is used as reference, we can achieve purity of up to $\sim94$ \% for the largest cylinder radius. Moreover, the purity of the sample increases when a larger cylinder is used, which is expected because it is less likely to misidentify a very bright nearby satellite as a central when a larger cylinder is used that is more likely to also contain the central galaxy. 

We also confirmed the purity of our sample in low-redshift galaxies ($0.1<z<0.3$) of the GAMA sample, where the satellite fraction remains high, around $\sim27$ \% (Fig. \ref{fig:satfrac}), because the algorithm might under-perform in this satellite-rich redshift space. We find, however, that the purity of the central sample is higher in this regime, which makes us confident that our central sample selection is reliable. 

Results from applying the algorithm to the MICE2 mock galaxy catalogue are also shown in Fig. \ref{fig:purcom}. The purity values that we achieve are very similar to the values we obtained using GAMA, except when the largest cylinder radius is used. This means that for the largest radius, the actual purity of our central sample is higher than the one we measure using GAMA. 

In addition, we tried to optimise our central selection using the stellar masses in the mock galaxy catalogues. First, we modified the algorithm so as to select the most massive galaxy in the cylinder centre instead of the brightest one. For this, we used the stellar mass present in the MICE catalogues, and plot the purity in Fig. \ref{fig:purcom}. The performance is worse than when the apparent brightness is used, suggesting that the central galaxy is more often the brightest galaxy in the halo, but not the most massive in terms of stellar mass. 

Lastly, instead of using a fixed cylinder radius to search for overdense regions, we used a per-galaxy cylinder radius, tied to the $R_{200}$ of the galaxy. To compute this, we used the stellar-to-halo mass relation computed for GAMA central galaxies \citep{shmf},
\begin{equation}
M_*^{\rm c}(M_{\rm h}) = M_{*,0}\frac{(M_{\rm h}/M_{\rm h,1})\,^{\beta_1}}{[1+(M_{\rm h}/M_{\rm h,1})]\,^{\beta_1-\beta_2}}\,,
\label{eq:shmf}
\end{equation}
where $M_*^{\rm c}$ is the stellar mass of the central galaxy and $M_{\rm h}$ the halo mass. We used the best-fit values from \citet{shmf} for the remaining parameters in this model and solved numerically for $M_{\rm h}$. We then computed the $R_{200}$ from $M_{\rm h}\equiv4\pi(200\bar{\rho}_{\rm m})R_{200}^3/3$, where $\bar{\rho}_{\rm m}=8.74\times10^{10}h^2$M$_\odot/$Mpc$^3$ is the comoving matter density. Fig. \ref{fig:purcom} shows that the purity of the sample generally increases when a more per-galaxy optimised cylinder is used.

It is clear that increasing the cylinder radius increases the purity of our central galaxy sample, but this comes at a cost. Specifically, the completeness of the sample drops as the radius increases, and we finally have fewer galaxies for our analysis. This is expected because larger cylinders encompass ever more central galaxies, making the sample less complete. Even the gain using the galaxy $R_{200}$ as a cylinder radius causes the completeness to drop by $\sim3$ \%. It is therefore important to find a good compromise between the sample purity and the total number of central galaxies that are used. 

\begin{figure}
        \resizebox{\hsize}{!}{\includegraphics{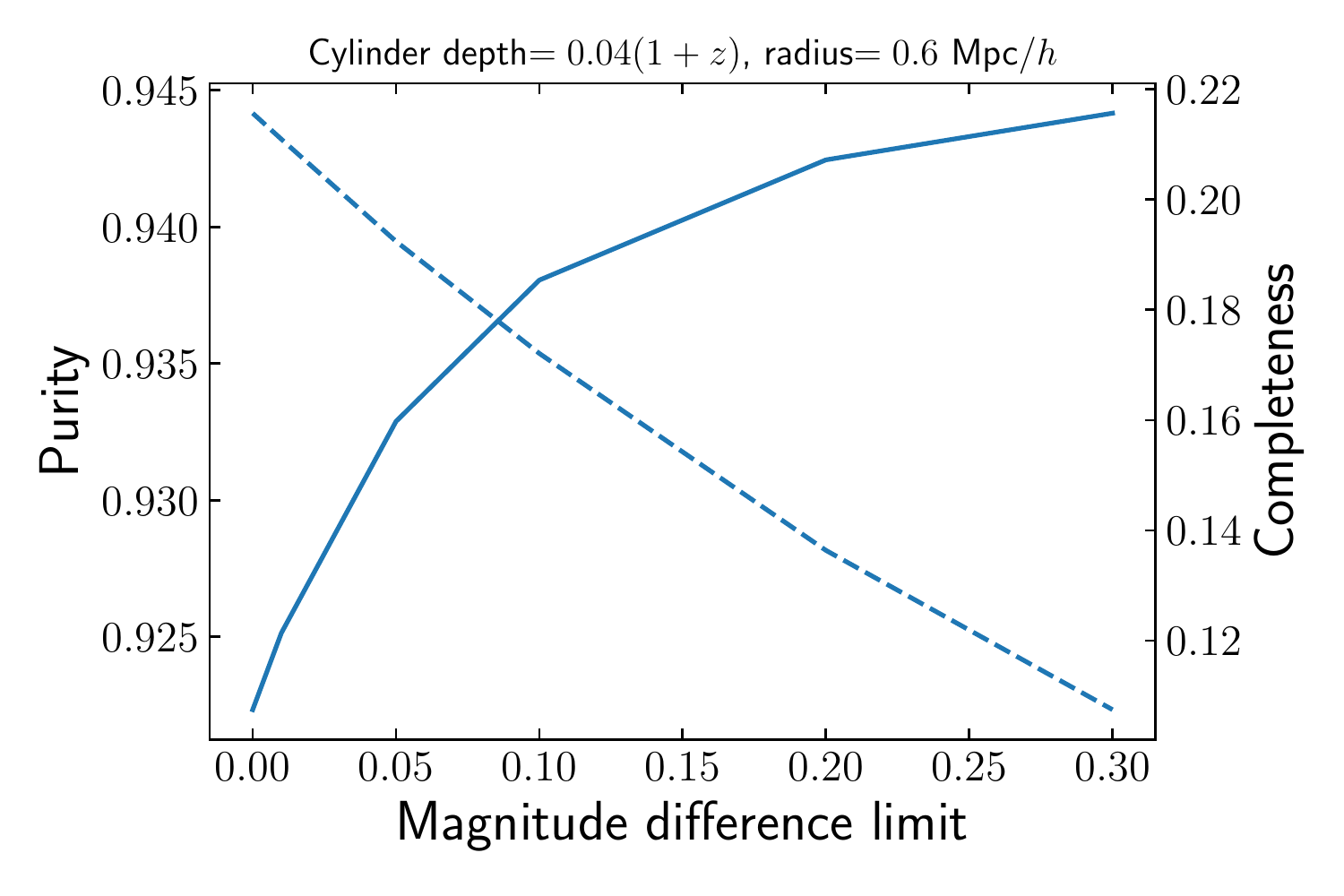}}
        \caption{Purity (solid, left $y$-axis) and completeness (dashed, right $y$-axis) of our central galaxy sample after rejecting centrals with a galaxy brighter than a magnitude difference from the brightness of the central galaxy, shown on the $x$-axis. The cylinder was fixed at 0.6 Mpc$/h$. Lines connect the individual points.}
        \label{fig:purcomDm}
\end{figure}

As a last step in this direction, we considered the second brightest galaxy in the cylinder and the difference in magnitude from the brightest galaxy. When two galaxies are in the same overdensity but are too close in magnitude, it is possible that the centre of the halo does not correspond to the brightest galaxy. Therefore we rejected centrals that have a galaxy inside the same cylinder up to a magnitude difference limit. We plot the purity of the central sample following this procedure as a function of the magnitude difference limit for a fixed cylinder of 0.6 Mpc$/h$ radius in Fig. \ref{fig:purcomDm}. The purity increases as the magnitude difference limit increases, but the completeness drops. 

Based on this, we chose to use a magnitude difference limit of 0.1 in our final sample. To increase the sample size without compromising much on its purity, we opted to use a fixed cylinder radius of 0.6 Mpc$/h$. With this setup, we achieve a purity of 93.4\%, as quantified from the overlap with the GAMA group catalogue. The total number of central galaxies for the whole KiDS-1000 area after masking is 138,607. Shape measurements are successfully obtained for 115,930 galaxies using a weight function with a scale equal to $r_{\rm iso}$ and 117,601 galaxies using $1.5r_{\rm iso}$. 

\subsection{Scaling with photo-$z$ accuracy}

Interestingly, the purity of the sample seems to plateau for large cylinders. To understand this better, we repeated the analysis using the mock galaxy catalogues and sampled photometric redshifts with three different values of accuracy, $\mathrm{d}z=\{0.02, 0.01, 0.0035\}(1+z)$. The first choice represents our lens galaxy sample, the second corresponds to the photometric redshifts achievable for luminous red galaxies \citep[LRGs,][]{Rozo2016,MJ}, and the last is the expected redshift accuracy from a narrow-band based survey, such as the Physics of the Accelerated Universe \citep[PAUS][]{PAUSphotozs}. 

\begin{figure}
        \resizebox{\hsize}{!}{\includegraphics{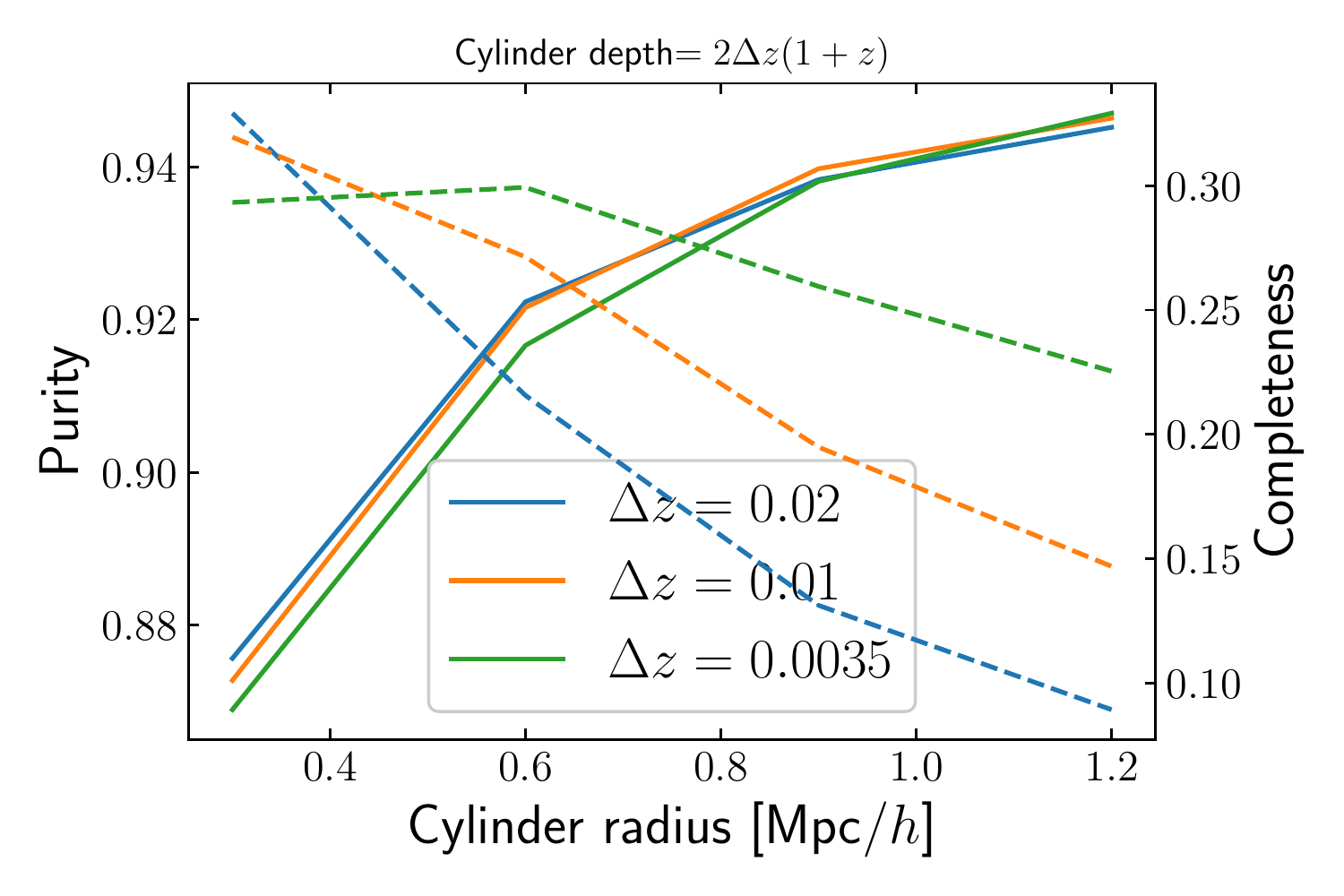}}
        \caption{Purity (solid lines) of our central galaxy sample selection as a function of the fixed cylinder radius. Lines connect the individual points. Results shown for three different simulated photometric redshift accuracies. The depth of the cylinder is equal to $\pm2$ times the redshift uncertainty. The completeness of the central galaxy sample is also shown on the right $y$-axis, overplotted with dashed lines. }
        \label{fig:purcomDz}
\end{figure}

The results are shown in Fig. \ref{fig:purcomDz}, where we plot the purity and completeness for the different redshift accuracies. We also changed the size of the cylinder in redshift space according to the redshift accuracy. Interestingly, the purity of the sample remains roughly the same in all three cases. As the redshift accuracy and cylinder depth decrease, the completeness of the sample increases as well. From this we conclude that improvements to the purity cannot be made by reducing the redshift uncertainty. 

The lack of improvement in purity can be understood that in lower mass groups the central halo galaxy does not always correspond to the brightest galaxy \citep[see e.g.][]{fBCG}. Increasing the redshift accuracy allows a better determination of centrals in less massive halos, which are expected to carry a weaker signal of halo ellipticity, however. Therefore it is better to increase the area of the survey, if possible, instead of the redshift accuracy. This justifies our choice of using the much larger area KiDS-1000 data and not the spectroscopic redshifts of the GAMA survey for our analysis.

\subsection{Sample characteristics}

We present here the characteristics of the final sample of central galaxies we compiled. The sample properties were obtained for the overlap of our KiDS-1000 sample with the GAMA survey, where an extensive photometry and stellar mass catalogue was used \citep[\texttt{StellarMassesLambdarv20,}][]{StellarMasses, Lambdar}. This catalogue provides estimates of the stellar mass, absolute magnitudes, and restframe colours of galaxies using fits to galaxy SEDs from photometry in the optical+NIR broad bands. 

In addition, we split the central sample into intrinsically red and blue galaxies. To do so, we isolated the red-sequence galaxies by inspecting the distribution of apparent $g-i$ colour versus $m_r$ in ten linear redshift bins in the redshift range of the lens sample. With this division, we obtained 62426 red and 53504 blue-lens galaxy sub-samples. Their average ellipticity modulo is the same as for the full sample, but their distributions show that slightly more blue galaxies have ellipticities with absolute values below $0.1$ or above $0.3$ than red galaxies.

\begin{figure}
        \resizebox{\hsize}{!}{\includegraphics{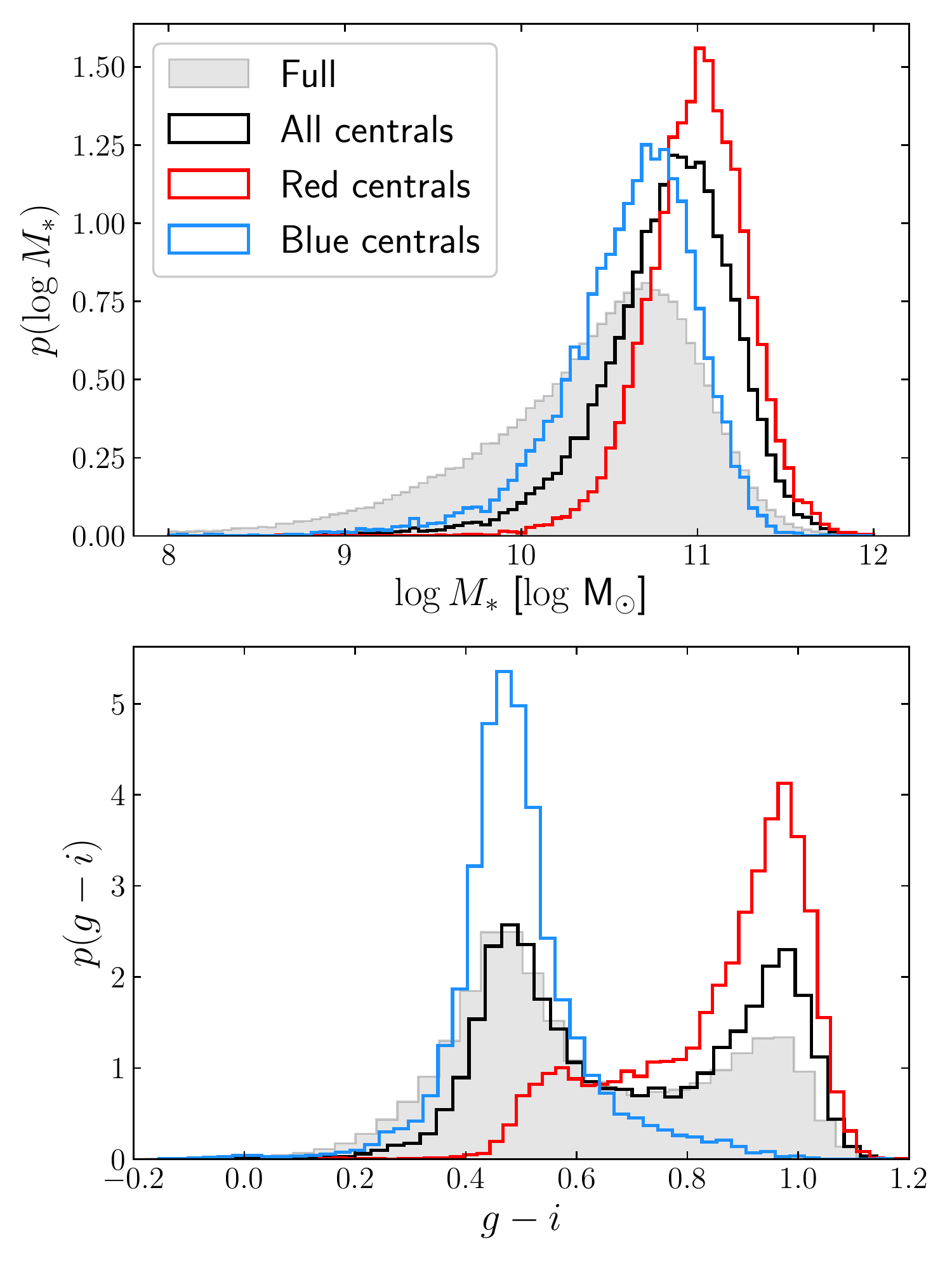}}
        \caption{Normalised distributions of the full sample (in filled grey) and our central galaxy sample (in black, red, and blue for all, red, and blue centrals, respectively) in the GAMA overlap. \emph{Top:} Distribution of stellar mass. \emph{Bottom:} Distribution of restframe dust corrected $g-i$ colour.}
        \label{fig:Mrcolorhist}
\end{figure}

In the top panel of Fig. \ref{fig:Mrcolorhist} we show the distribution of stellar mass for the full galaxy population in the bright KiDS-1000 sample (restricted to the overlap with GAMA), as well as for our central galaxy sample, also divided into red and blue centrals. The central galaxies are generally more massive than galaxies in the whole population, as expected. The mean stellar mass of the red and blue central sample is $\sim10^{11}$ M$_\odot$ and $10^{10.6}$ M$_\odot$, respectively.

In addition to this, we show the distribution of restframe $g-i$ colours, corrected for dust extinction, in the bottom panel of Fig. \ref{fig:Mrcolorhist}, again for the full KiDS-1000 and central (all, blue, and red) galaxy sample. The central galaxy sample consists of generally more red galaxies than the full sample. The colour distributions of our selection of red and blue centrals generally also follows the expected restframe $g-i$ distribution, which makes us confident in our colour selection. We note that a small number of relatively blue galaxies enter our red-galaxy sample, which is an effect of our imperfect colour split based on photometric redshift data and a visual inspection. However, given the number of these galaxies, we do not expect a cleaner sample selection to alter our results. 

\section{Method}
\label{sec:Methodology}

Gravitational lensing has the effect of coherently distorting light rays of background galaxies (sources) from the intervening matter along the line of sight. Because galaxies are biased tracers of the matter density in the Universe, we expect to find a correlation between the position of foreground galaxies (lenses) and the shapes of the source galaxies. In its weak regime, the effect is very small, and the observed ellipticities of source galaxies are only affected by about 1\%. Large statistical ensembles of lens-source galaxy pairs are therefore required to extract the weak-lensing signal. 

For lens and source galaxies, we used the third flattening, $\epsilon=\epsilon_1+\rm{i}\epsilon_2$, as an ellipticity measure, which is related to the ratio of semi-minor to semi-major axis, $q$, by $|\epsilon|=(1-q)/(1+q)$. We can then express the tangential and cross ellipticity of source galaxies with respect to the lens position as
\begin{align}
\epsilon_+&=-\epsilon_1\cos(2\theta)-\epsilon_2\sin(2\theta)\,,\label{eq:tangential_ellipticity}\\
\epsilon_\times&=\epsilon_1\sin(2\theta)-\epsilon_2\cos(2\theta)\,,
\label{eq:cross_ellipticity}
\end{align}
where $\theta$ is the position angle of the line connecting the lens-source galaxy pair. When averaged over pairs, $\epsilon_+$ provides an unbiased but noisy estimate of the gravitational shear $\gamma$, that is, $\langle \epsilon_+\rangle\approx\gamma_+$, which can then be related to the excess surface mass density through
\begin{equation}
\Delta\Sigma(R)=\bar{\Sigma}(<R)-\Sigma(R) =\gamma_+(R)\Sigma_{\rm crit}\,,
\label{eq:DeltaSigmaGamma}
\end{equation}
with $\Sigma_{\rm crit}$ the critical surface density, defined by
\begin{equation}
\Sigma_{\rm crit}=\frac{c^2}{4\pi G}\,\frac{D_{\rm s}}{D_{\rm l} D_{\rm ls}}\,.
\label{eq:Sigma_crit}
\end{equation}
In this equation, $c$ and $G$ are the speed of light and gravitational constant, respectively, $D_{\rm s}$ is the angular diameter distance to the source galaxy, $D_{\rm l}$ to the lens galaxy and $D_{\rm ls}$ between the lens and source galaxy. Equation \eqref{eq:DeltaSigmaGamma} holds true only when an azumuthally averaged ensemble of lenses is considered. 

The isotropic (azimuthally averaged) part of the lensing signal can be calculated from the data using the estimator
\begin{equation}
\widehat{\Delta\Sigma}=\left(\frac{\sum_{\rm ls} w_{\rm ls}\epsilon_+\Sigma_{\rm crit}}{\sum_{\rm ls}w_{\rm ls}}\right)\,,
\label{eq:DeltaSigma}
\end{equation}
where the sum runs over all lens-source galaxy pairs that fall in a given projected radius bin. We weighted each pair with ellipticity weights, $w_{\rm s}$, computed by \emph{lens}fit, which accounts for the uncertainty in the shear estimate, and define
\begin{equation}
w_{\rm ls}=w_{\rm s}\Sigma_{\rm crit}^{-2}\,.
\end{equation}
Because galaxy redshifts are computed through photometry, it is important to account for the full posterior redshift distribution of source galaxies, $p(z_{\rm s})$ (see Sect. \ref{sec:KiDS}) when computing $\Sigma_{\rm crit}$. This is done with equation
\begin{equation}
\Sigma_{\rm crit}^{-1}=\frac{4\pi G}{c^2}\int_{z_{\rm l}}^{\infty}\frac{D_{\rm l}(z_{\rm l}) D_{\rm ls}(z_{\rm l}, z_{\rm s})}{D_{\rm s}(z_{\rm s})}p(z_{\rm s})\mathrm{d}z_{\rm s}\,.
\end{equation}

\subsection{Anisotropic lensing model}

We modelled the anisotropic part of the lensing signal following the formalism presented in \citet{Tim2015}, which is based on work by \citet{Natarajan} and \citet{Mandelbaum2006}. The excess surface mass density of a lens is modelled as
\begin{equation}
\Delta\Sigma_{\rm model}(r,\Delta\theta) = \Delta\Sigma_{\rm iso}(r)[1+4f_{\rm rel}(r)|\epsilon_{h,a}|\cos(2\Delta\theta)]\,.
\label{eq:lensmodel}
\end{equation}
Here, $\Delta\Sigma_{\rm iso}$ is the excess surface mass density for a spherical halo (estimated from data using Eq. \eqref{eq:DeltaSigma}) and $\Delta\theta$ is the position angle coordinate in the lens plane, measured from the halo semi-major axis. The ellipticity of the halo is probed by the galaxy ellipticity, therefore we are sensitive only to the aligned component of the halo ellipticity with the galaxy, $|\epsilon_{h,a}|$ The anisotropy of the elliptical halo lensing is described by $f_{\rm rel}(r)$, which depends on the assumed halo density profile and is generally a function of the projected separation $r$. For elliptical halos that are not described by a single power-law, $f_{\rm rel}(r)$ needs to be computed numerically \cite[see e.g.][]{Mandelbaum2006}, and we interpolated this quantity (using a cubic interpolation) from tabulated values. In order to avoid systematic biases in our anisotropic lensing signal measurement, it is also necessary to define the excess surface mass with lens and source ellipticities rotated by $\pi/4$, where we have
\begin{equation}
\Delta\Sigma_{\rm 45,model}(r,\Delta\theta)=\Delta\Sigma_{\rm iso}(r)[4f_{\rm rel,45}(r)|\epsilon_{h,a}|\cos(2\Delta\theta+\pi/2)]\,,
\label{eq:lens45model}
\end{equation}
where $f_{\rm rel,45}(r)$ is obtained in the same manner as $f_{\rm rel}(r)$. 

The quantity of interest is the ratio of the halo ellipticity modulo to the galaxy ellipticity modulo, $\tilde{f}_{\rm h}=|\epsilon_h|/|\epsilon_g|$. However, we can only measure this quantity weighted by the average mis-alignment angle between the halo and host galaxy semi-major axis, $\Delta\phi_{\rm h,g}$. Consequently, the measured quantity $f_{\rm h}=\tilde{f}_{\rm h}\langle\cos(2\Delta\phi_{\rm h,g})\rangle$ (where we also assume that the misalignment angle does not depend on $|\epsilon_{\rm h}|$). 
In order to extract $f_{\rm h}$ from data, we used the following estimators:
\begin{equation}
\widehat{f\Delta\Sigma}=\frac{\sum_{\rm ls} w_{\rm ls}\epsilon_+\Sigma_{\rm crit} |\epsilon_{\rm l}|\cos(2\phi_{\rm ls})}{\sum_{\rm ls}w_{\rm ls}|\epsilon_{\rm l}|^2\cos^2(2\phi_{\rm ls})}\,,
\label{eq:fDS}
\end{equation}
and 
\begin{equation}
\widehat{f_{45}\Delta\Sigma}=-\frac{\sum_{\rm ls} w_{\rm ls}\epsilon_\times\Sigma_{\rm crit} |\epsilon_{\rm l}|\sin(2\phi_{\rm ls})}{\sum_{\rm ls}w_{\rm ls}|\epsilon_{\rm l}|^2\sin^2(2\phi_{\rm ls})}\,,
\label{eq:f45DS}
\end{equation}
where $\phi_{\rm ls}$ is the angle between the lens semi-major axis and the position vector connecting the lens-source galaxy pair. These two estimators can be predicted from $f_{\rm h}f_{\rm rel}\Delta\Sigma_{\rm iso}$ and $f_{\rm h}f_{\rm rel,45}\Delta\Sigma_{\rm iso}$, respectively. However, the estimators are easily contaminated by systematic errors in the lensing signal measurements, such as imperfections due to incorrect PSF modelling or cosmic shear from structures between the lens and the observer \citep{Mandelbaum2006, Tim2015}. An estimator that is insensitive to these systematic effects can be constructed by subtracting the two,
\begin{equation}
\widehat{(f-f_{45})\Delta\Sigma} = \widehat{f\Delta\Sigma}-\widehat{f_{45}\Delta\Sigma}\,.
\label{eq:finalestimator}
\end{equation}
To measure the ellipticity ratio, we used this estimator. The analysis we followed is described below.

\subsection{Extracting $f_h$}
\label{sec:freqanalysis}

To measure $f_{\rm h}$ from data, we considered the two estimators $x_i=\widehat{\Delta\Sigma}_i$ and $y_i=\widehat{(f-f_{45})\Delta\Sigma}_i/(f_{\rm rel}(r_i)-f_{\rm ref,45}(r_i))$, where the index $i$ runs over the radial bins over which we calculated the lensing signal, and $r_i$ is the central value of that bin. These are two random Gaussian variables, which prevents us from simply computing their fraction $m=y_i/x_i$, which would lead to a biased estimate of $f_{\rm h}$. To overcome this, we considered for a given $m$ the quantity $y_i-mx_i$. This is a random Gaussian variable drawn from $N(0,w_i^{-1})$, with $w_i^{-1}=\sigma_y^2+m^2\sigma_x^2$ and $\sigma_x,\sigma_y$ are the error on the measured estimators $x_i$ and $y_i$, respectively \citep{Mandelbaum2006}. 

The following sum ratio
\begin{equation}
\frac{\sum_i w_i (y_i-mx_i)}{\sum_i w_i}\sim N\left(0,\frac{1}{\sum_i w_i}\right)
\end{equation}
is also a random Gaussian variable. Based on this, we determined confidence intervals of $\pm1\sigma$ in measuring $m$ by considering the inequality
\begin{equation}
\frac{-Z}{\sqrt{\sum_i w_i}}<\frac{\sum_i w_i (y_i-mx_i)}{\sum_i w_i}< \frac{+Z}{\sqrt{\sum_i w_i}}\,.
\label{eq:freqapproach}
\end{equation}
We used $m$ drawn from a grid and calculated $f_{\rm h}$ by requiring $f_{\rm h}=m(Z=0)$. We also determined the $\pm1\sigma$ intervals by setting $Z=\pm1$. In addition, we computed the reduced $\chi^2$ from $n$ radial bins using 
\begin{equation}
\chi_{\rm red}^2=\frac{\sum_i w_i(y_i-m(Z=0)x_i)^2}{n-1}\,.
\label{eq:chisquare}
\end{equation}
This method does not take into account the off-diagonal elements of the covariance matrix of our measurements. These where estimated to be very small (see Sect. \ref{sec:Ellipticity}), however, with the standard deviation of the correlation matrix off-diagonal elements being $4\times10^{-2}$.

\section{Halo ellipticity}
\label{sec:Ellipticity}

\begin{table*}
        \caption{Results from fits to the weak-lensing signal for the full sample and for the red and blue central galaxy sub-samples. The mean stellar mass is shown for galaxies in the KiDS-1000 and GAMA overlap, quoted from the \texttt{StellarMassesLambdarv20} catalogue. We also show the best fit and error of $M_{200}$ from the NFW profile fits to the isotropic weak-lensing signal and the resulting ellipticity ratio $f_{\rm h}$ fit, with its reduced $\chi^2$, according to Sect. \ref{sec:freqanalysis}, obtained using the two different weight function sizes to measure lens galaxy shapes (see Sect. \ref{sec:KiDS}).}
        \label{tab:fh}
        \centering
        \renewcommand{\arraystretch}{1.5}
        \begin{tabular}{l c c c c c c}
                \hline\hline
                Sample & $M_*$ [$10^{10}$ M$_\odot$] & $M_{200}$ [$10^{12}$ M$_\odot$] & $f_{\rm h}$ $(r_{\rm wf}/r_{\rm iso}=1)$ & $\chi^2_{\rm red}$ $(r_{\rm wf}/r_{\rm iso}=1)$ & $f_{\rm h}$ $(r_{\rm wf}/r_{\rm iso}=1.5)$ & $\chi^2_{\rm red}$ $(r_{\rm wf}/r_{\rm iso}=1.5)$  \\
                \hline
                Full & 9.28 & $3.74\pm0.16$ & $0.27^{+0.19}_{-0.18}$ & $1.14$ & $0.50\pm0.20$ & $1.41$ \\
                Red & 12.14 & $6.69\pm0.30$ & $0.34\pm0.17$ & $0.80$ & $0.55\pm0.19$ & $0.90$ \\
                Blue & 5.63 & $1.35\pm0.14$ & $0.08^\pm0.53$ & $0.83$ & $0.28\pm0.55$ & $1.23$ \\
                \hline
        \end{tabular}
\end{table*}

\begin{figure*}
        \centering
        \includegraphics[width=17cm]{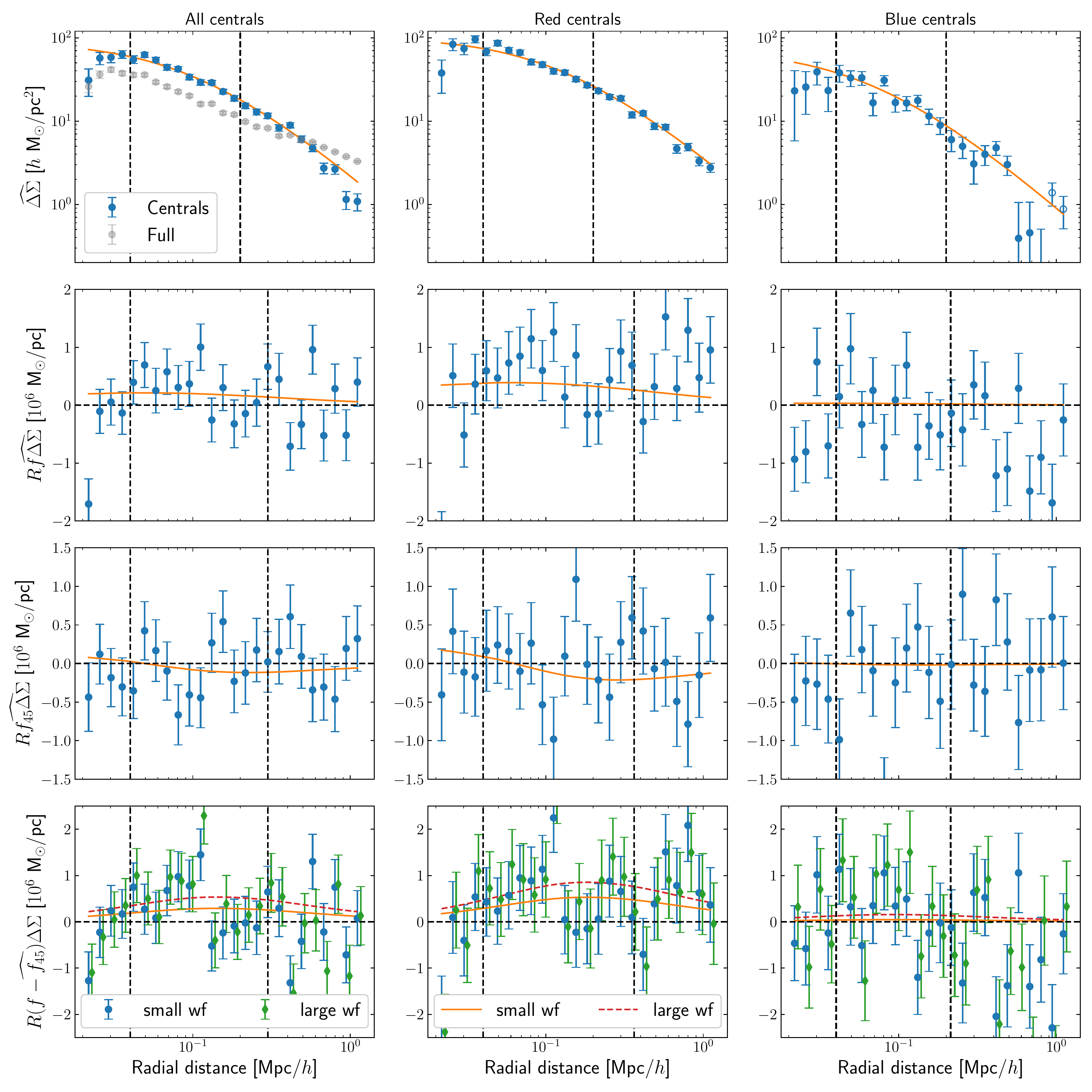}
        \caption{Measurements of the weak-lensing signal around our central galaxy sample. The first column shows results obtained for all centrals, and the second and third columns show results for the red and blue sub-samples, respectively (with open circles placed for negative measurements). The first row shows the isotropic lensing signal, the second and third row show the anisotropic lensing signal obtained with the estimators of Eqs. \eqref{eq:fDS}-\eqref{eq:f45DS}, and the last row shows the difference, Eq. \eqref{eq:finalestimator}. The best-fit NFW profile is overplotted in the first row, with dashed vertical lines depicting the ranges that were used during the fit. We also show the isotropic lensing signal of the full KiDS-1000 sample as grey points as a comparison to the signal obtained using only the central galaxies. For the next rows we show the best-fit NFW profile multiplied by the best-fit $f_{\rm h}$ and $f_{\rm rel}, f_{\rm rel,45}$ and their difference, respectively, as well as the ranges we used during the fit with dashed lines. In the last row, we show with green diamonds and the dashed red line the data and model with the best-fit $f_{\rm h}$ obtained using a larger weight function of $r_{\rm wf}=1.5r_{\rm iso}$ to measure the lens galaxy shapes.}
        \label{fig:measurements}
\end{figure*}

We measured the weak-lensing signal around our central galaxy sample using 25 radial bins, logarithmically spaced between 20 kpc$/h$ and 1.2 Mpc$/h$. We restricted the sample to lens galaxies with well-defined ellipticities, $0.05<\epsilon_{\rm l}<0.95$. The median redshift of the lenses is 0.26, and their average ellipticity is 0.188 for shapes obtained using weight function of $r_{\rm iso}$ and 0.183 when using $1.5r_{\rm iso}$. We fitted the isotropic weak-lensing signal with a Navarro–Frenk–White (NFW) profile \citep{NFW,NFWequations} and fixed the concentration-mass relation according to \citet{Duffy}, 
\begin{equation}
c=5.71\left(\frac{M_{200}}{2\times10^{12}\mathrm{M}_\odot/h}\right)^{-0.084}(1+z)^{-0.47}\,,
\label{eq:Duffy}
\end{equation}
to finally obtain an estimate for the scale radius $r_s$. This was then used to calculate $f_{\rm rel}(r)$ and $f_{\rm rel,45}(r)$. We note that our measurements of $f_{\rm h}$ are not very sensitive to the concentration-mass relation. Changing the constant in Eq. \eqref{eq:Duffy} by 20\% shifts our measured value by at most $\sim0.3\sigma$, therefore we did not consider a more complicated relation. In NFW profile fits, we used the mean redshift of the full lens sample. We also restricted the fit to the range from 40 kpc$/h$ up to 200 kpc$/h$. The first limit minimises signal from baryons in the centre of the halo and contamination of the source galaxy shear by the extended light of each lens \citep{Tim2015,Sifon2018}. The upper limit ensures that we do not include contributions from the two-halo term when fitting the lensing signal. This is a conservative limit because we do not expect a strong two-halo term in our lensing signal because our galaxy sample is only very little contaminated by satellite galaxies.

To calculate the covariance of our measurements, we used a bootstrap technique. We sampled $10^5$ random bootstrap samples from the lens catalogue (with replacement) and used this data vector to calculate the covariance matrix, obtaining error bars for our measurements from its diagonal elements. This technique ignores errors due to sample variance from the large-scale structure. However, these are expected to be negligible given the scales we probe. We tested this by computing the covariance and errors from a per-area bootstrap technique, dividing the survey into 1 deg$^2$ patches and computing the lensing signal in each patch. We then selected $10^5$ random bootstrap patches, weighting them by the number of lenses (because patches with significantly fewer than average lenses will have a more uncertain signal measurement) and derived fully consistent error bars. We also find that the off-diagonal elements of the covariance matrix are negligible at all scales, justifying the analysis outlined in Sect. \ref{sec:freqanalysis}.

We present our measurements in Fig. \ref{fig:measurements} for the case when the shape of the lens galaxy is measured using a weight function with scale equal to $r_{\rm iso}$. Results for the full, red, and blue lens galaxy sample are shown in the left, right, and middle column, respectively. The first row shows the isotropic lensing signal measurement, spherically averaged, as well as the best-fit NFW profile, with the ranges used in the fit indicated with dashed vertical lines. We also overplot the isotropic lensing signal obtained from the full KiDS-1000 bright-end catalogue in the same redshift range in the top left panel for comparison. Our central galaxy sample is generally more massive and is not affected by a strong two-halo term, in contrast to the full sample. The resulting average halo mass for the three sub-samples is listed in Table \ref{tab:fh}. We also verified that $\Delta\Sigma_{45}$, which is calculated by substituting the tangential with the cross ellipticity component in Eq. \eqref{eq:DeltaSigma}, is consistent with zero at all measured scales. This is expected because a spherically averaged cross component is not generated by gravitational lensing, and it serves as a useful sanity check for a potential systematic offset.

In the next three rows of Fig. \ref{fig:measurements} we present the measurement of the anisotropic lensing signal for the three sub-samples. We used these measurements to calculate the ellipticity ratio, $f_{\rm h}$, following Sect. \ref{sec:freqanalysis}, as well as the 1$\sigma$ confidence intervals. For the $f_{\rm h}$ measurement, we used scales from 40 kpc$/h$ up to the estimated $r_{200}$ for the corresponding galaxy sample, which are shown as dashed lines in the figure. For visualisation, we overplot the best-fit NFW profile of the corresponding galaxy sample, multiplied by $f_{\rm rel}, f_{\rm rel,45}$ or their difference, accordingly, and the best-fit value of $f_{\rm h}$. 

The resulting values of $f_{\rm h}$ as well as the reduced $\chi^2$ of the fit are presented in Table \ref{tab:fh}. We see that the ellipticity ratio is fitted reasonably well, as expressed by the $\chi^2$ values. For the full sample we measure an ellipticity ratio of 0.27 with a 1.5$\sigma$ statistical significance. For the red galaxies, the measured ratio is higher, 0.34, and the significance also increases to 2-$\sigma$. Finally, we do not measure a significant ellipticity ratio for blue galaxies.

\subsection{Misalignment dependence on galaxy scale}

Following the results presented in the previous section, we remeasured the ellipticity ratio, $f_{\rm h}$, using lens galaxy shapes with a weight function of scale equal to $1.5r_{\rm iso}$. When a larger weight function is used, the measured shapes are more sensitive to the morphology of outer galaxy regions. The mean ellipticity of the lens sample is measured to be very similar when using the two weight functions (with a difference of 0.005), and their distributions were inspected to be nearly the same. Therefore any difference measured in $f_{\rm h}$ will be directly related to differences in the mean misalignment angle, $\langle \cos(2\Delta\phi_{\rm h,g})\rangle$.

The isotropic lensing signal obtained with the larger weight function is statistically the same because the lensing sample is not systematically different. We show the anisotropic lensing measurements in the last row of Figure \ref{fig:measurements} with green diamonds and the model obtained using the best-fit $f_{\rm h}$ for a large weight function with a dashed red line (see also Table \ref{tab:fh}). The measured signal for $(f-f_{45})\Delta\Sigma$ is higher, although only at a level of $\sim1\sigma$. When the full sample was analysed, we measure an $f_{\rm h}=0.50\pm0.20$, which is $\sim1.5$ times higher than the value obtained using the smaller weight function. For the red and blue galaxy sub-samples, we find $f_{\rm h}=0.55\pm0.19$ and $f_{\rm h}=0.28\pm0.55$, respectively, both of which are higher than the values obtained by using a smaller weight function. For red galaxies the detection of a non-zero ellipticity ratio is increased to $2.9\sigma$, while blue galaxies are still found to have a value $f_{\rm h}$ fully consistent with zero.

Based on this analysis, we suggest that outer galaxy regions are more aligned with the shape of the dark matter halo. This agrees with other observations, where central galaxies were found to be more aligned with their satellite galaxy distributions when the shape measurement used was more sensitive to their outer regions \citep{Huang1, Georgiou2}. The physical processes causing this behaviour can be either tidal interactions between the central galaxy and the dark matter halo affecting the outer, less strongly bound galaxy regions more strongly, or it might be that infalling material to the central galaxy generally follows the ellipticity of the dark matter halo.

\subsection{Comparison with the literature}

Our analysis closely follows work done in previous studies. \citet{Mandelbaum2006} used a very similar estimator on a much larger lens sample, split into colour and luminosity. For their L6 luminosity bin, which is closer to the mean luminosity of our sample, they found $f_{\rm h}=0.29\pm0.12$ for red and $f_{\rm h}=1.0^{+1.3}_{-0.9}$ for blue galaxies, but we note that a sign inconsistency in their model computation might have affected these results \citep{Tim2015}. \citet{Edo2012} also studied a large lens sample consisting of less massive galaxies than ours, and found $f_{\rm h}=0.19\pm0.10,0.13\pm0.15,$ and $-0.16^{+0.18}_{-0.19}$ for all, red, and blue lens samples, respectively. Following the same method, \citet{Tim2015} studied a sample of lenses split into colour and stellar mass, and found $f_{\rm h}=-0.04\pm0.25$ for all red lenses and $f_{\rm h}=0.69^{+0.37}_{-0.36}$ for all blue ones. They also provided predictions of $f_{\rm h}$ from the Millennium Simulation \citep{Millennium}, which agrees with the values we obtain here. The studies above used an almost identical method as we did and lens samples much larger than ours, but the samples were likely contaminated by satellite galaxies. Our study indicates the importance of selecting central galaxies for an anisotropic lensing signal measurement. 

The pioneering works of \citet{Hoekstra2004} and \citet{Parker} conducted similar measurements of $f_{\rm h}$ for single-band photometric data and found $f_{\rm h}=0.77^{+0.18}_{-0.21}$ and $f_{\rm h}=0.76\pm0.10$, respectively. However, these results were not corrected for the spurious signal introduced by other effects that align lens and source ellipticities. This may have biased
the resulting fh measurements to high values \citep{Tim2015}.

Focussing on the brightest group galaxies (BGG) of the GAMA group catalogue specifically (using groups with more than five members), \citet{Edo2017} detected an halo ellipticity of $\epsilon_{\rm h}=0.38\pm0.12$ using the BGG semi-major axis as a proxy for the halo orientation and focussing on scales below 250 kpc. Similar ellipticity has also been detected for dark matter halos of galaxy clusters \citep{Evans&Bridle, Clampitt, SDSSclusters, Clash}. For comparison, we find $\epsilon_{\rm h}=0.051^{0.036}_{-0.034}$ for the full sample and $\epsilon_{\rm h}=0.064\pm0.032$ for red galaxies using the average lens galaxy ellipticity of our sample and assuming zero misalignment angle.

The galaxies in these groups and clusters are generally more massive than our central galaxy sample (see Table \ref{tab:fh}), with the GAMA BGG sample having a mean stellar mass of $2.25\times10^{11}$ M$_\odot$, and cluster central galaxies are typically more massive than that. In order to confirm whether more massive galaxies in our sample have an higher ellipticity ratio, we selected galaxies based on stellar mass, obtained running the code Le Phare \citep{LePhare} on the KiDS-1000 nine-band photometry (Wright et al. in prep.). Using all galaxies with $M_*>1.58\times10^{11}$ M$_\odot$ , we find $f_{\rm h}=0.24\pm0.19$, which is slightly higher than the value for the whole sample.

The ellipticity we obtain is significantly lower than what is measured in galaxy groups. This suggests either that halos of galaxy groups and clusters are more elliptical than those of relatively isolated galaxies\footnote{Our sample consists of both BGG and field galaxies, the latter expected to be either isolated galaxies or BGGs of groups whose satellites are too faint to be detected within the imposed magnitude limit.} or that the mean misalignment between halos and galaxies is smaller for group and cluster central galaxies. In cosmological simulations, higher mass halos where found to be more elliptical and less misaligned with their host galaxy than lower mass halos, which agrees with the trend observed here \citep[e.g.][]{Tenneti2014,Marco2015,Elisa2017}. However, we did not measure a significant increase in $f_{\rm h}$ when we restricted our sample to high stellar mass galaxies, which leaves the interpretation unclear. 

Another possible reason for the discrepancy may be differences in the shape measurement of the lenses. Shapes of lens galaxies were derived using a generally large weight function in \citet[][and private communication]{Edo2017}. We measure a larger $f_{\rm h}$ when using a larger weight function for measuring shapes of lens galaxies in our sample, which might explain at least part of the low-halo ellipticity value we find in comparison to galaxy groups.

\section{Conclusions}
\label{sec:Conclusions}

We measured the anisotropic lensing signal and halo-to-galaxy ellipticity ratio of galaxies for a bright sample ($m_r\lesssim20$) with accurate redshifts acquired through a machine-learning technique, trained on a similar spectroscopic sample (GAMA, $m_{r,\rm{petro}}<19.8$). We minimised satellite contamination because it would complicate the interpretation and modelling of the measured signal. To construct the sample, we identified galaxies in regions of high galaxy number density and selected the brightest galaxy (in $r$ band) in a cylindrical area. We assessed the purity of our central galaxy sample using the overlap with GAMA and found the purity to be $=93.4\%$. Similar values were obtained using the MICE mock galaxy catalogues (built from N-body cosmological simulations). 

We used the central galaxy sample as lens galaxies and background sources from the KiDS-1000 shear catalogues. We also split the lens sample into intrinsically red and blue galaxies. Using the measured lensing signal, we extracted the ellipticity ratio $f_{\rm h}$ (weighted by the misalignment angle between the galaxy and the halo semi-major axis) using an estimator that was not affected by systematic errors such as incorrect PSF modelling and cosmic shear. We measure $f_{\rm h}=0.27^{+0.19}_{-0.18}$ for the full sample and $f_{\rm h}=0.34\pm0.17$ for an intrinsically red sub-sample, respectively, while for blue galaxies the ratio is fully consistent with zero. Our measurements agree with predictions based on cosmological simulations, and we demonstrated the importance of using a highly pure sample of central galaxies to measure the halo ellipticity.

Our results generally agree with studies of similar galaxy samples. However, we find a significantly lower halo ellipticity when we compared our results to central galaxies of galaxy groups and clusters. Cosmological simulations predict that lower mass halos are rounder and/or more strongly misaligned with their host halo than more massive halos, which may explain part of this difference. Using shape estimates that are more sensitive to outer galaxy regions, we find a higher value for $f_{\rm h}$, specifically, $0.5\pm0.2$ and $0.55\pm0.19$ for the full and red sample, respectively, rejecting the hypothesis of round halos and/or randomly aligned galaxies with respect to their parent halo at $2.5$ and $2.9\sigma$. This suggests there is a galaxy-scale dependence of the misalignment angle $\Delta\phi_{\rm h,g}$, with outer regions of the host galaxy being more aligned with its dark matter halo.

Our results can also be connected with the difference found between the predicted galaxy intrinsic alignment signal of dark matter halos and the observationally measured alignment of galaxies, which are found to have a much lower signal \citep[e.g.][]{MSalignment,okumura}. In addition, galaxy intrinsic alignments have been observed to depend on the galaxy properties, with more luminous (and therefore massive) galaxies indicating a stronger alignment amplitude than less luminous ones \citep[][]{Singh, Harry}. This trend of the alignment signal points in the same direction as the decreasing misalignment of halos and galaxies with increasing halo mass seen here and in cosmological simulations.

\bibliographystyle{aa} 
\bibliography{references} 

\begin{acknowledgements}
        The author list is divided into three tiers; the first three authors co-wrote this paper, MB provided infrastructure data and made a significant informative contribution to the paper, and the rest of the authors contributed infrastructure and are listed alphabetically.

        We acknowledge funding from the European Research Council through grant numbers 770935 (AD, HHi, AW) and 647112 (CH, BG). HHo and AK acknowledges support from Vici grant 639.043.512, financed by the Netherlands Organisation for Scientific Research (NWO). JTAdJ is supported by the NWO through grant 621.016.402. KK acknowledges support by the Alexander von Humboldt Foundation, the Royal Society and Imperial College. MB is supported by the Polish Ministry of Science and Higher Education through grant DIR/WK/2018/12, and by the Polish National Science Center through grants no. 2018/30/E/ST9/00698 and 2018/31/G/ST9/03388. HHi is supported by a Heisenberg grant of the Deutsche Forschungsgemeinschaft (Hi 1495/5-1).  CH acknowledges support from the Max Planck Society and the Alexander von Humboldt Foundation in the framework of the Max Planck-Humboldt Research Award endowed by the Federal Ministry of Education and Research. HYS acknowledges the support from NSFC of China under grant 11973070, the Shanghai Committee of Science and Technology grant No.19ZR1466600 and Key Research Program of Frontier Sciences, CAS, Grant No. ZDBS-LY-7013.
        
        Based on data products from observations made with ESO Telescopes at the La Silla Paranal Observatory under programme IDs 177.A-3016, 177.A-3017 and 177.A-3018, and on data products produced by Target/OmegaCEN, INAF-OACN, INAF-OAPD and the KiDS production team, on behalf of the KiDS consortium. OmegaCEN and the KiDS production team acknowledge support by NOVA and NWO-M grants. Members of INAF-OAPD and INAF-OACN also acknowledge the support from the Department of Physics \& Astronomy of the University of Padova, and of the Department of Physics of Univ. Federico II (Naples). 
        
        GAMA is a joint European-Australasian project based around a spectroscopic campaign using the Anglo-Australian Telescope. The GAMA input catalogue is based on data taken from the Sloan Digital Sky Survey and the UKIRT Infrared Deep Sky Survey. Complementary imaging of the GAMA regions is being obtained by a number of independent survey programmes including GALEX MIS, VST KiDS, VISTA VIKING, WISE, Herschel-ATLAS, GMRT and ASKAP providing UV to radio coverage. GAMA is funded by the STFC (UK), the ARC (Australia), the AAO, and the participating institutions.
        
        This work has made use of CosmoHub.     CosmoHub has been developed by the Port d'Informació Científica (PIC), maintained through a collaboration of the Institut de Física d'Altes Energies (IFAE) and the Centro de Investigaciones Energéticas, Medioambientales y Tecnológicas (CIEMAT), and was partially funded by the "Plan Estatal de Investigación Científica y Técnica y de Innovación" program of the Spanish government.

\end{acknowledgements}

\end{document}